\documentclass[draft]{agujournal2019}
\usepackage{url} 
\usepackage{lineno}
\usepackage{amsmath}
\usepackage{amsfonts}
\usepackage{enumitem}

\usepackage{soul}
\draftfalse

\journalname{Journal of Advances in Modeling Earth Systems (JAMES)}
\graphicspath{{}{./}}

\begin{document}

\title{Using a rare event sampling technique to quantify extreme El Niño event statistics}

\authors{Sarah Packman\affil{1}, Justin Finkel\affil{2}, Dorian S. Abbot\affil{2}, Eli Tziperman\affil{3}}

\affiliation{1}{Department of Physical Oceanography, MIT-WHOI Joint Program in Oceanography / Applied Ocean Science and Engineering, Woods Hole, MA, USA}
\affiliation{2}{Department of the Geophysical Sciences, University of Chicago, Chicago, IL, USA}
\affiliation{3}{Department of Earth and Planetary Sciences and School of Engineering and Applied Sciences, Harvard University, Cambridge, MA, USA}

\correspondingauthor{Sarah Packman}{sarah.packman@whoi.edu}


\begin{keypoints} 
\item Extreme El Niño events are rare; simulating a significant number to allow their study requires prohibitively long model integrations

\item We apply TEAMS, a rare event sampling method, to simulate such rare events at a lower cost, allowing to learn their statistics and dynamics

\item As a proof of concept, TEAMS produced rare event statistics in the Zebiak-Cane ENSO model at 20$\%$ of the cost of a direct numerical simulation
\end{keypoints}

\begin{abstract}
Extreme El Niño events, such as occurred in 1997--1998, can induce severe weather on a global scale, with significant socioeconomic impacts that motivate efforts to understand them better. However, extreme El Niño events are rare, and even in a very long direct numerical simulation (DNS) occur too infrequently for robust statistical characterization. This study seeks to generate extreme El Niño event model data at a lower cost, while preserving statistical fidelity, using a rare event sampling technique, which preferentially devotes computational resources toward extreme events by generating a large, branched ensemble of interrelated trajectories through successive targeted perturbations. We specifically use the ``trying-early adaptive multi-level splitting'' (TEAMS) algorithm, which is well-suited for El Ni{\~n}o's relative timescales of predictability and event duration.  We apply TEAMS to the Zebiak-Cane model, an intermediate-complexity ENSO model for which it is feasible to run a long DNS (500,000 years) for validation. We compare extreme El Niño event return time estimates from TEAMS to those from the long DNS to assess TEAMS’ accuracy and efficiency. We find that TEAMS accurately reproduces the return time estimates of the DNS at about one fifth the computational cost. Therefore, TEAMS is an efficient approach to study rare ENSO events that can be plausibly applied to full-complexity climate models.
\end{abstract}

\section*{Plain Language Summary}
Extreme El Ni{\~n}o events, such as occurred in 1997--1998, are rare and therefore difficult to simulate without running models for extremely long durations, which is computationally prohibitive when using state-of-the-art global climate models. We simulate extreme El Ni{\~n}o events more efficiently using a rare event sampling technique, which preferentially devotes computational resources toward simulating extreme events while preserving statistical accuracy. Rare event sampling techniques involve branching one simulation off of another (“splitting”) to generate successively more extreme iterations of an event. We apply the algorithm to an intermediate-complexity El Nino model for which it is feasible to run a long direct numerical simulation (DNS) to use as “ground truth.” We compare extreme event statistics calculated from the algorithm to those calculated from the long DNS, finding that the rare event sampling approach accurately reproduces the results at one fifth the computational cost. Therefore, this approach to studying rare El Ni{\~n}o events can likely be applied to more realistic climate models. 

\section{Introduction}

The large El Niño of 1997--1998 led to extreme weather on a global scale, including drought, forest fires, and heatwaves in some parts of the world and intensified rainfall and flooding in others \cite{McPhaden-Zebiak-Glantz-2006:enso}. Ecological impacts included widespread mortality of reef-building corals \cite{Wilkinson-2004:status}. Modeling extreme El Niño events remains a challenge due to the computational costs associated with generating enough model realizations of an extreme event (which by definition appears infrequently) to calculate meaningful statistics. This paper explores the application of the rare event sampling technique ``trying-early adaptive multi-level splitting'' \cite<TEAMS,>{Finkel-OGorman-2024:bringing} to the generation of model data of rare extreme El Niño events. We use the Zebiak-Cane ENSO model \cite{Zebiak-Cane-1987:model} as a test case for the algorithm's accuracy and efficiency in the context of ENSO, and study its potential to be applied to expensive state-of-the-art global climate models.

Extreme El Niño events are characterized by high sea surface temperatures (SSTs) throughout the equatorial Pacific as quantified by a large NINO3 index: the running monthly average of SSTs in the region of the eastern equatorial Pacific spanning $\pm$5 \textdegree N, 210--270 \textdegree E \cite<for example,>{McPhaden-1999:genesis, Cai-Borlace-Lengaigne-et-al-2014:increasing}. Extreme El Niño events are expected to double in frequency in the period 1991--2090 as compared to the baseline period 1890--1991, according to a multi-model analysis by \cite{Cai-Borlace-Lengaigne-et-al-2014:increasing} with 20 Coupled Model Inter-Comparison Project models, each run for 200 years. Proposed mechanisms for the increased frequency include stronger coupling between the ocean and atmosphere in the regime of warmer SSTs (intensified Bjerknes feedback) and an overall reduction in meridional and zonal SST gradients \cite{Cai-Borlace-Lengaigne-et-al-2014:increasing}. However, details about the most extreme El Niño events, such as the incremental dropoff in frequency with increasing peak SST anomaly, remain difficult to discern for extreme and therefore rare events because running complex global climate models (GCMs) over long periods is prohibitively expensive.

Rare event sampling techniques seek to devote the bulk of computational resources toward modeling the tail end of the distribution of possible events \cite{Bucklew-2004:importance}. A major advantage of rare event sampling as opposed to purely statistical extrapolation \cite<e.g., Extreme Value Theory;>{Coles-2001:introduction} is that sampling provides the dynamical evolution leading up to the events, allowing one to study their physical mechanisms. Often, such algorithms involve building an ensemble by creating multiple copies (with small perturbations applied, or ``splitting'') of a simulation that captures the event of interest. This ensemble, which contains many distinct instances of an extreme event, can then be used to calculate rare event statistics with lower variance than when using direct numerical simulation (DNS) approaches. These techniques were originally developed to calculate the probability of particle transmission in nuclear reactions \cite{Kahn-Harris-1951:estimation}, but have since been used to calculate rare event probabilities in a wide variety of fields, including biophysics \cite{Zuckerman-Chong-2017:weighted}, planetary dynamics \cite<destabilization of Mercury's orbit,>{Abbot-Webber-Hadden-et-al-2021:rare}, natural resource management
\cite{Jansen-Mandjes-Taimre-2023:rare}, and weather and climate modeling \cite{Ragone-Wouters-Bouchet-2018:computation, Plotkin-Webber-ONeill-et-al-2019:maximizing, Webber-Plotkin-ONeill-et-al-2019:practical, Lucente-Herbert-Bouchet-2022:committor, Finkel-OGorman-2024:bringing, Noyelle2025evolution,lancelin2025ai}.

There are several distinct flavors of rare event sampling methods, with different advantages and drawbacks. Diffusion Monte Carlo (DMC) operates by perturbing simulations at a sequence of chronological times, and is well suited for amplifying long-lasting, slow-moving events, such as hot seasons \cite{Ragone-Wouters-Bouchet-2018:computation} and wet seasons \cite{Wouters-Schiemann-Shaffrey-2023:rare}. A variant based on quantiles (QDMC) generalizes the method to more suddenly-onsetting events (where``sudden'' is relative to the model timestep length) like tropical cyclones \cite{Webber-Plotkin-ONeill-et-al-2019:practical} and planetary orbit destabilization \cite{Abbot-Webber-Hadden-et-al-2021:rare}. This is done by regularizing the ensemble with quantile remapping, as well as making use of skillful predictors, which give a strong advance signal of which ensemble members promise to become most extreme \emph{in the future} of a simulation. There is in fact an optimal predictive variable (the ``committor function''), which has been calculated in great detail in a handful of reduced-order climate models \cite<>[the latter on a simple 3-variable ODE ENSO model]{Finkel-Abbot-Weare-2020:path, Finkel-Webber-Gerber-et-al-2021:learning, Jacques-Dumas-Westen-Bouchet-et-al-2022:data, Lucente-Herbert-Bouchet-2022:committor} but is generally intractable. In many cases of practical interest there is no known predictor skillful enough at a long enough range for perturbations to take effect---at least on a standard DMC schedule. The splitting algorithm described in the next section adapts the perturbation schedule to address this situation.

\paragraph*{Trying-Early Adaptive Multi-Level Splitting.} In this study, we  use ``trying-early adaptive multi-level splitting'' \cite<TEAMS;>{Finkel-OGorman-2024:bringing,Finkel-OGorman-2025:rare}, a variation of adaptive multi-level splitting \cite<AMS;>{Lestang-Ragone-Bréhier-et-al-2018:computing,Cerou2019adaptive}. AMS begins with an initial ensemble of multiple trajectories sampled from different initial conditions across the steady-state distribution (in practice, sampled from a short DNS). Then, similarly to QDMC, AMS selects some less-extreme trajectories to kill and some more-extreme trajectories to split at the onset time of the event of interest. The selection step is stochastic, but follows precise rules that associate probability weights to each trajectory. The end result of many iterations of splitting/ killing is an ensemble of branched trajectories that simulate increasingly extreme events, which, together with their probability weights, can then be used to calculate extreme event statistics. TEAMS is distinct from AMS in that the splitting step occurs some time \textit{before} the extreme event of interest. As a result, the new split-off trajectory has the potential to surpass the old trajectory in severity. TEAMS is an appropriate choice of rare event sampling technique for studying El Niño because it is meant to simulate events that have a shorter duration than an ensemble's \emph{dispersion timescale}: the time it takes for trajectories to spread apart following small perturbations \cite{Finkel-OGorman-2024:bringing}. The timing of the splitting step in TEAMS is crucial to generate multiple, diverse realizations of an extreme event. 

Our goal is to develop the needed methodology to simulate extreme El Niño events using TEAMS. Rather than applying TEAMs to a full-complexity GCM, we begin by using the Zebiak-Cane \cite<ZC,>{Zebiak-Cane-1987:model} model as an intermediate-complexity test case. The ZC model has been used extensively to study aspects of the ENSO cycle \cite{Zebiak-Cane-1987:model, Clement-Seager-Cane-2000:suppression, Tziperman-Cane-Zebiak-1995:irregularity, Tziperman-Zebiak-Cane-1997:mechanisms, Mann-Cane-Zebiak-et-al-2005:volcanic, Cane-2005:evolution} and successfully forecast it \cite{Cai-Kalnay-Toth-2003:bred, Xie-Jin-2018:two, Geng-Jin-2022:enso}. This study thus serves as a proof of concept, illustrating that TEAMs has the potential to study extreme ENSO events in full-complexity GCMs.

\section{Methods}
\label{sec:methods}

\subsection{The Zebiak-Cane Model}

The atmospheric component of the Zebiak-Cane model \cite{Zebiak-Cane-1987:model} is essentially a steady-state linearized atmosphere driven by convective heating calculated based on the wind divergence and SST. The oceanic component is a mixed-layer/thermocline shallow water model and spans 30\textdegree S--30\textdegree N (2\textdegree\ meridional resolution) and 129.375--320.625 \textdegree E (5.625\textdegree\ zonal resolution). Simulations run with this model setup produce irregular chaotic oscillations with periods of 2--7 years \cite{Tziperman-Cane-Zebiak-1995:irregularity}. The ZC model effectively reproduces the basic features of the ENSO system, and while it cannot be considered a realistic model of El Niño, it is appropriate for the purposes of this study given its computational efficiency and plausible representation of El Niño physical mechanisms.

\begin{figure}[!tbp]
   \begin{center}
   \includegraphics[width=0.75\textwidth]{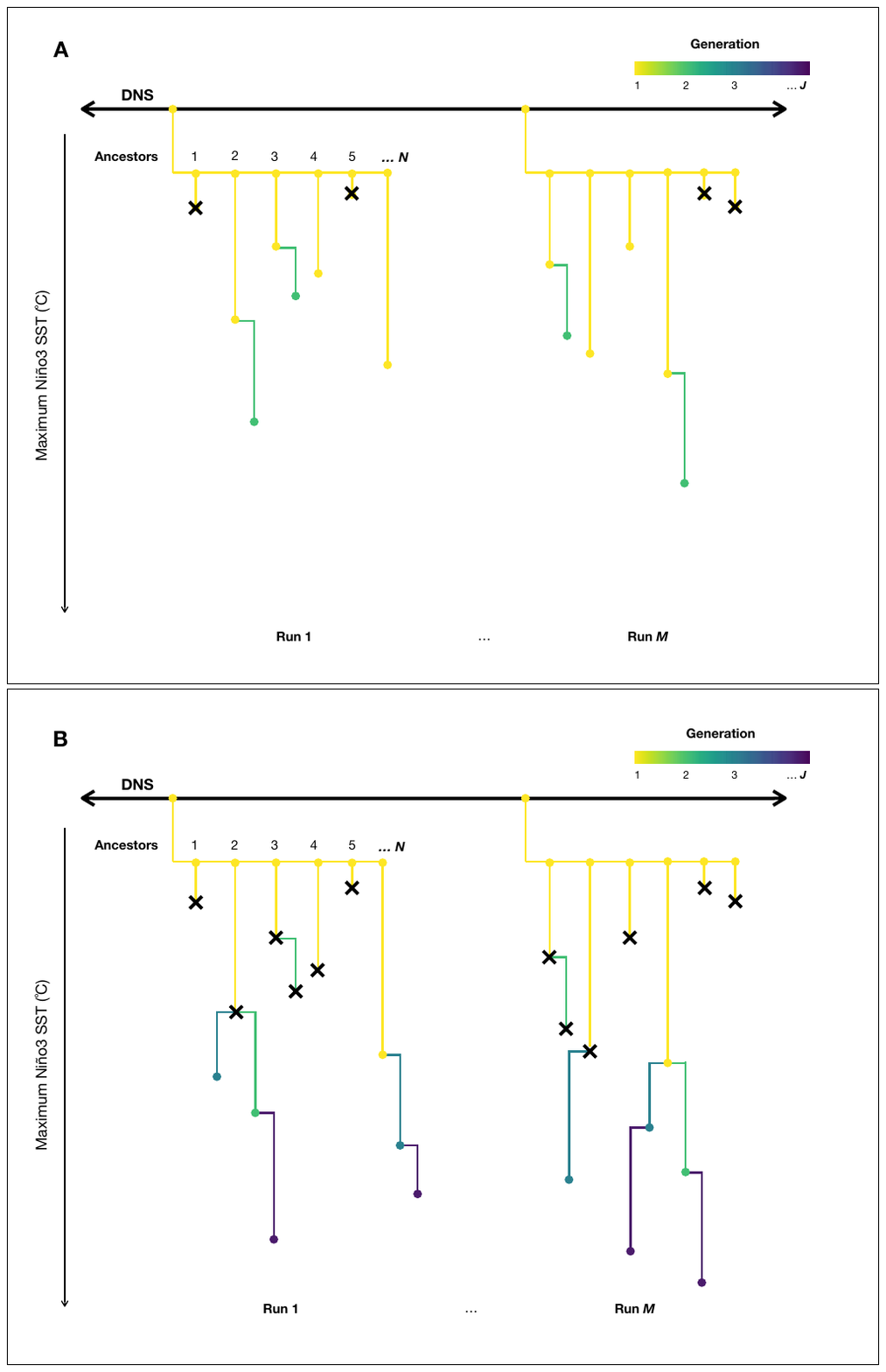}
   \end{center}
   \caption{Schematic overview of a TEAMS ensemble. Thick, black line across the top represents a short DNS. Yellow nodes on the DNS represent the beginnings of distinct runs of TEAMS. (A) At the start of each run, $N$ ancestors (yellow offshoots) of duration $T$ are created by taking an initial condition from a DNS, applying a small perturbation, and running the simulation forward for $N \times T$ years. The $k$ lowest-scoring ancestors (lowest maximum NINO3 SST indices, y-axis) are killed (black X's), and $k$ new offspring are produced (green offshoots). (B) Subsequent TEAMS generations. In each generation, the threshold NINO3 level $\ell$ is raised ($\ell_1<\ell_2<\ell_3<...<\ell_J$), the $k$ ensemble members with the lowest scores are killed, and $k$ offspring are produced from randomly selected parents.}
   \label{fig:TEAMS_family_tree_schematic}
\end{figure}

\subsection{TEAMS}

Our implementation of TEAMS proceeds as follows, similarly to \cite{Finkel-OGorman-2025:rare}, with several tunable parameters whose values we report after the general description.
\begin{enumerate}[nosep]
  \item Begin with an ensemble of $N$ initial ``ancestors'', each $T$ years in length. The initial condition for the first ancestor is selected randomly from a direct numerical simulation (DNS), which is long compared to the ENSO cycle but short compared to the return periods we aim for. The initial condition is perturbed so that the simulation diverges from the DNS. The initial condition for the second ancestor is taken as the final state of the first, again with a random perturbation added, and so on for ancestors $3,...,N$ (Fig.~\ref{fig:TEAMS_family_tree_schematic}).
  \item Assign each ancestor an initial weight of $w_i=1$, and an initial state of ``alive'', meaning it can be used to produce other (``child'') simulations. The sum of the weights of all ensemble members is therefore $N$, a property that will be preserved as the algorithm proceeds even as the number of members increases. These weights are used later to estimate without bias the rare event statistics, in particular the PDF of the NINO3 index.
  \item Define the ``score'' of each live member as the maximum value of the NINO3 index over its length-$T$ years timespan. Arrange the live members in increasing order of score, and kill those with the $k$ lowest scores. These dead members will henceforth spawn no additional descendants, but they will contribute to statistical estimation through their (dead) weights, which henceforth remain fixed. Denote by $W_{\text{live}}$ the sum of live weights. 
  \item Take the average of the $k$ and $k-1$ lowest scores as the threshold $\ell_1$ for this iteration of TEAMS. The scores of all live ensemble members, by construction, exceed this threshold. 
  \item Randomly select a simulation from the live ensemble members to serve as a new parent.
  \item Find when the parent simulation crosses the threshold ($t = t_{\text{cross}}$).
  \item Define new ``split'' initial conditions for ``child'' simulations from the parent $\delta$ years before this time step ($t_{\text{split}} = t_{\text{cross}} - \delta$) and apply a spatially white random noise $W(x,y)$ to the SST field of these initial conditions. Run a child simulation from these initial conditions from $t=t_{\text{split}}$ to $t = T$ years, and increment the total running cost by $T-t_{\text{split}}$ years.
  \item Determine if the child simulation exceeds the threshold $\ell_1$ for this iteration.
  \begin{enumerate}
    \item If the child exceeds the threshold (``success''): mark it as alive (able to produce future offspring), and assign it the same weight $w$ as its parent. Adjust all live weights (including that of the new child) by a multiplicative factor $W_{\text{live}}/(W_{\text{live}}+w)$ to preserve the total sum of (alive+dead) weights as $N$. 
    \item If the child does not exceed the threshold (``failure''): Discard the child trajectory. That is, set its weight to zero, so it will not contribute to statistical estimates. Go back to step 6 and randomly select a new parent. 
  \end{enumerate}
  \item Repeat steps 6--8 until $k$ new successful children have been produced, or until hitting the computational budget (see step 11 below).
  \item Perform repeated level raisings: repeat steps 3--9 until the total simulation cost exceeds the prescribed maximum allowed cost of $T_\text{max}$ years. This completes a full TEAMS run.
  \item At any point, if the total cost (duration) of simulations exceeds a predetermined threshold of $T_\text{max}$ years, terminate the process and archive the resulting ensemble as a completed ``run'' of TEAMS.
  \item Repeat TEAMS (steps 1--11) $M$ times, starting from distinct initial conditions and using different seeds for random number generation. Note that each TEAMS run is completely decoupled from the others, so the $M$ runs are fully parallelizable. Although each individual run separately gives unbiased statistical estimates, multiple runs together can be used to both quantify statistical uncertainty and reduce it.
\end{enumerate}

\subsection{Parameter Selection} 

TEAMS' efficiency is measured by the cost (number of simulation years) required to reproduce the return time statistics of a long DNS run with a similar statistical error estimate. Multiple parameters interact to affect the accuracy and uncertainty estimate in TEAMS: a larger $N$ (number of ancestors) reduces run-to-run variance by sampling the attractor more completely; a larger $M$ (more runs) reduces overall variance at the standard rate of $1/M$; a longer advance split time $\delta$ increases statistical independence from parent to child, thus decreasing variance, but eventually at the expense of not being able to affect the peak magnitude in the case of sensitivity to initial conditions due to chaotic behavior or stochastic forcing. Experimentation is usually needed to select hyperparameters, which we did here as well. 

We settled on the following parameter values. The block size (integration length) is $T=25$ years, the number of TEAMS runs is $M=50$, the number of initial ancestors is $N=16$, the maximum computational budget per run of TEAMS is $T_\text{max} = 2000$ years, and the discard rate (fraction of ensemble members ``killed'' after each level raising) is 0.25, so that $k=0.25\times N=4$. 

\begin{figure}[tbp!]
   \begin{center}
    \includegraphics[width=0.6\textwidth]{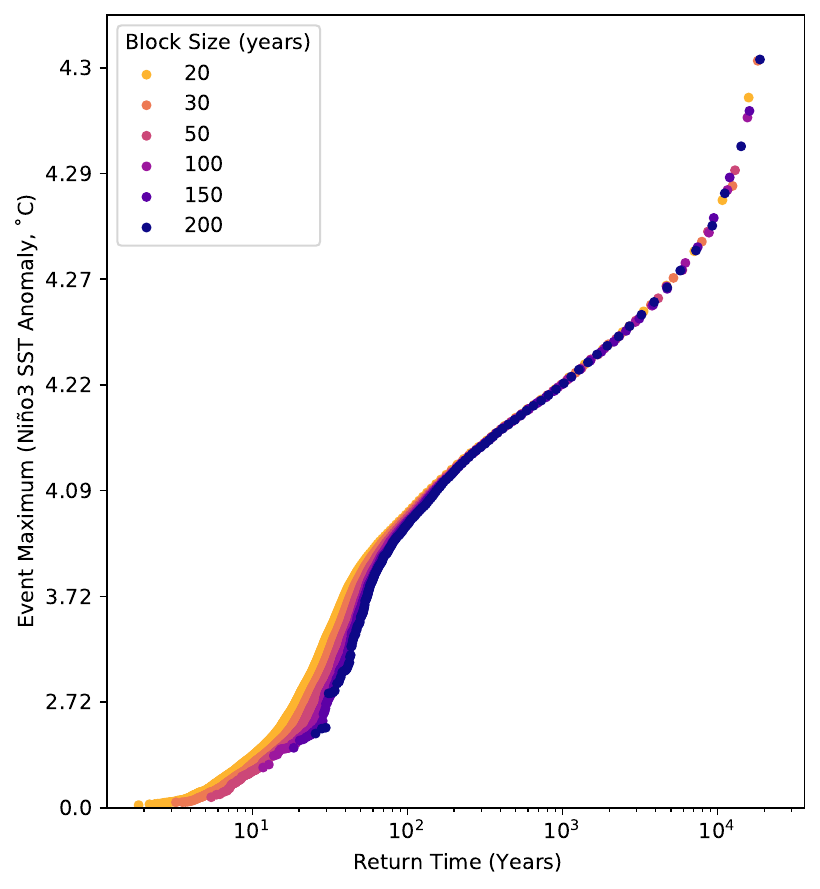}
   \end{center}
   \caption{Return level as a function of return time, calculated from a DNS (500,000 years) using the modified block maxima method (Poisson approximation) for a range of block sizes (20, 30, 50, 100, 150, 200). The return level $y$ axis is stretched as $-\log(1-\text{NINO3 SST} / 4.3$ \textdegree C), which serves to highlight the tail end of the distribution of NINO3 SST maxima. The figure shows that all block sizes result in similar return time estimates for periods that are larger than the block size.}
   \label{fig:block-sizes-comparison}
\end{figure}

We experimented with different block sizes from 20 to 100 years and settled on $T = 25$, finding it short enough to likely contain at most one extreme event, but long enough that consecutive block maxima are statistically independent, as required by the Poisson process approximation described in Section~\ref{sec:methods}\ref{sec:return-time-calculations}. Fig.~\ref{fig:block-sizes-comparison} demonstrates that lengthening the time horizon further beyond 25 years does not change return time estimates for the most extreme events from the DNS (model ground truth). We chose $N=16$ because it is found empirically to be the minimum number of ancestors required to achieve a spread in block maxima, ensuring sufficient ensemble diversity \cite{Finkel-OGorman-2025:rare}.  

The amplitude of the spatially white noise added to initial conditions is uniformly distributed in $[-\sigma,\sigma]$ where $\sigma=0.1$ \textdegree C. This noise level was chosen to be small enough to avoid an unphysical initial state and for the child to inherit ``extremity'' from its parent, but large enough to substantially alter the trajectory of the child over its integration time.

Next, to set the advance split time $\delta$, consider the running average root mean square error (RMSE) calculated for each of the ensembles as,
\begin{align}
    \text{RMSE}(t) = \sqrt{\frac{1}{t\, K}\sum_{t'=1}^t{}\sum_{i=1}^K\|T_{\text{DNS}}(t') - T_{i}(t')\|^2},
    \label{eq:RMSE}
\end{align}
where $T_{\text{DNS}}(t')$ is a vector of the $6 \times 12$ grid of SST anomaly in the NINO3 area from each time step of the DNS, $T_i(t')$ represents the same from the $i$th TEAM ensemble member, and $t$ represents time. The RMSE was then normalized by the root-mean-square distance (RMSD) of the DNS, which is equivalent to $\lim_{t\to\infty}\text{RMSE}(t)$ and estimated as 
\begin{align}
    \text{RMSD}_{\text{DNS}} = \sqrt{\frac{1}{K_p}\sum_{i,j} \|T(t_i) - T(t_j)\|^2},
    \label{eq:RMSD}
\end{align}
where $T(t_i)$ and $T(t_j)$ are again vectors of the temperature in the NINO3 region, from randomly chosen time steps on the DNS, and $K_p$ is the total number of pairs in the DNS.

RMSE represents the typical distance between an ensemble member and the DNS as a function of time since the ensemble member's initial condition. Taking the running average reduces variance in the RMSE estimate for a given ensemble member, though at the cost of biasing RMSE at early times; we deem this an acceptable tradeoff given the heuristic nature of choosing an advance split time. RMSD represents the typical distance between any two points on the dynamical attractor at statistical steady state. Immediately following a split of a child run from a parent run, the ratio $\epsilon(t) = {\text{RMSE}(t)}/{\text{RMSD}}$ is close to zero but grows exponentially. Over time, as the ensemble members diverge, $\epsilon(t)$ approaches 1 and will continue to fluctuate about 1 due to finite ensemble size.

\begin{figure}[!tbp]
  \begin{center}
  \includegraphics[width=0.9\textwidth]{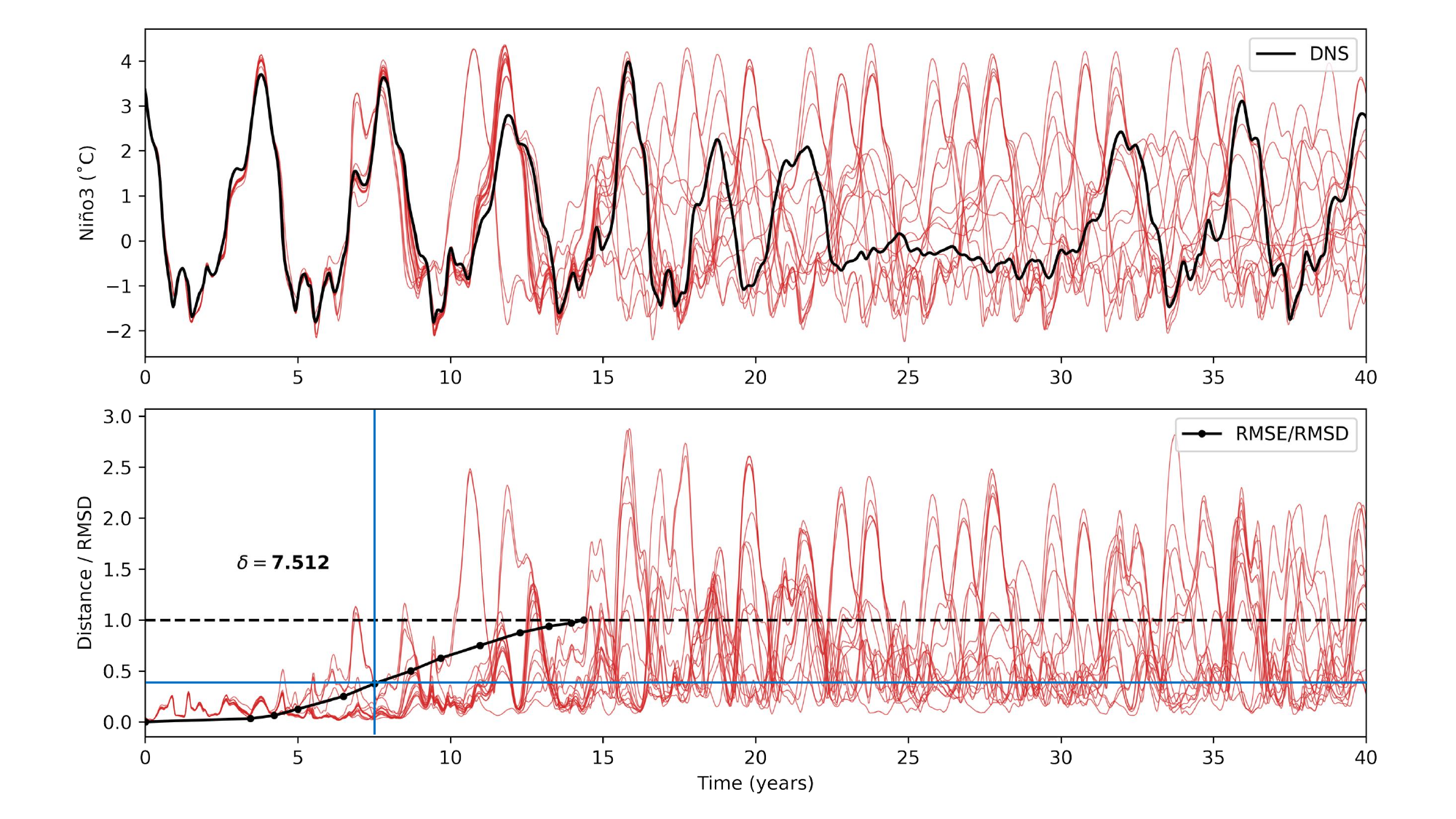} 
  \end{center}
  \caption{Quantification of advance split time ($\delta$) with a maximum perturbation amplitude $\sigma = 0.1$ \textdegree C. (A) Example divergence of NINO3 of child trajectories (thin red lines) from parent DNS (black line). (B) Example distances between child trajectory and parent DNS as a function of time (thin red lines). Black line: the ensemble-averaged  $\epsilon = \text{RMSE}(t) /\text{RMSD}$, which exceeds ${3}/{8}$ after 7.5 years (blue lines).}
  \label{fig:advance_split}
\end{figure}

We followed the rule of thumb from \cite{Finkel-OGorman-2024:bringing} to choose $\delta$ as the average time (over different initial conditions) for $\epsilon(t)$ to exceed ${3}/{8}$, leading to $\delta\approx7.5$ years (Fig.~\ref{fig:advance_split}). We deemed this choice appropriate, as it is long enough to allow the child to produce a substantially larger NINO3 maximum than its parent does, but short enough for that maximum to occur at a similar time. The value of 7.5 years makes intuitive physical sense for ENSO, as it is not much larger than the length of the typical four-year ENSO cycle. The predictability time scale of observed El Niño events is not well-known, yet may be of the order of 1--2 years, and may be different from that in the ZC model, implying that $\delta$ may need to be adjusted when TEAMS is applied to other models.

\begin{figure}[!t]
    \begin{center}
    \includegraphics[width=0.9\textwidth]{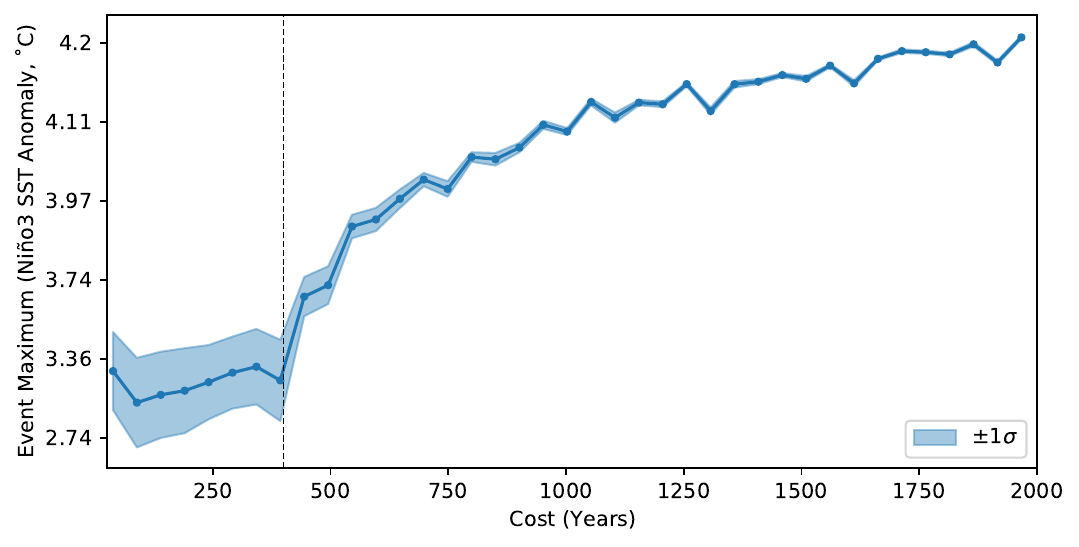}
    \end{center}
    \caption{Maximum achieved NINO3 SST anomaly averaged across $M=50$ runs of TEAMS as a function of the maximum allowed computational cost $T_{\text{max}}$ (total model years). Ancestor simulations (cost $<400$, left of black dashed line) have a range of scores; once level raisings begin (cost $>400$ years), scores increase and the error estimate (variance between TEAMS runs) decreases.}
    \label{fig:max_v_cost}
\end{figure}

We chose $T_{\text{max}}=2000$ years as the computational budget. Beyond this point, we find that there are negligible changes in the maximum NINO3 estimate (TEAMS' score) for higher computational cost (Fig.~\ref{fig:max_v_cost}). Next, we ensured that intermediate values of NINO3 were sufficiently represented in the early stages of level raising (i.e., that the jump in block maxima following the initiation of level raisings, seen as the vertical dashed line in Fig.~\ref{fig:max_v_cost}, is not too abrupt) by selecting a not-too-large discard rate of $k/N=0.25$. Finally, we increased the number of repetitions, $M$, until the variance in return time estimates calculated from TEAMS (described in Section~\ref{sec:methods}\ref{sec:return-time-calculations}) was comparable to that calculated from the long DNS, resulting in $M = 50$.

\subsection{Return Time Calculations}
\label{sec:return-time-calculations}

We follow \cite{Lestang-Ragone-Bréhier-et-al-2018:computing} and \cite{Finkel-OGorman-2024:bringing} in estimating the return times of extreme events. Given a set of ``blocks'', that is, contiguous segments of simulation with length $T$ either from DNS or from a branched ensemble, we can use the method of block maxima to estimate the return time $\tau(\ell)$ for temperature level, $\ell$, using the Poisson process approximation following \cite{Lestang-Ragone-Bréhier-et-al-2018:computing}:
\begin{align}
   \tau(\ell)=\frac{-T}{\log[1-p_T(\ell)]},
   \label{eq:block-max-return-time}
\end{align}
where $p_T(\ell) = 1-\text{CDF}(\ell)$ is the probability that NINO3 exceeds $\ell$ in a block of duration $T$. Given a DNS dataset, we split it into chunks of length $T$. Given a TEAMS dataset ($M$ runs with $N$ ancestors each, and possibly different total members), we take each member as a different block. To calculate the return times using TEAMS, we estimate the CDF at temperature $\theta$ as 
\begin{align}
    \mathrm{CDF}(\theta)=\frac1M\sum_{m=1}^M\frac1N\sum_{j=1}^{J_m}w_{m,j}\mathbb{I}\{\theta_k\leq\theta\},
\end{align}
where $J_m$ is the total number of members produced in the $m$th run of TEAMS (no greater than $T_\text{max}/T=2000/25=80$ in our case); $w_{m,j}$ is the weight of the $j$th member within the $m$th run, which when summed over $j$ yields $N$; and $\mathbb{I}\{\cdot\}$ is the indicator function: one if its argument is true, and otherwise zero. 

Thus, the smallest $\theta$ has the largest associated value $p_T$ (slightly smaller than 1). The ordered pairs $(\theta, p_T(\theta))$ can then be plugged into Eq.~\eqref{eq:block-max-return-time} to calculate $\tau(\theta)$.

The 95\% confidence intervals for TEAMS return times were calculated via 5,000 iterations of bootstrapping with replacement over the $M$ TEAMS runs. When calculating return time using the DNS, time blocks were treated interchangeably while bootstrapping.

\section{Results}
\label{sec:results}

TEAMS was able to reproduce the return time estimates of a long DNS with greater efficiency, using only 20\% of the CZ simulation years used in the DNS. In the three subsections below (~\ref{sec:TEAMS-accuracy},~\ref{sec:TEAMS-efficiency},~\ref{sec:successful-child}) we  consider in turn TEAMS' accuracy, efficiency, and dynamical output, exploring the development of an example successful splitting step.

\subsection{TEAMS Accuracy}
\label{sec:TEAMS-accuracy}

\begin{figure}[tbp!]
   \begin{center}
   \includegraphics[width=0.9\textwidth]{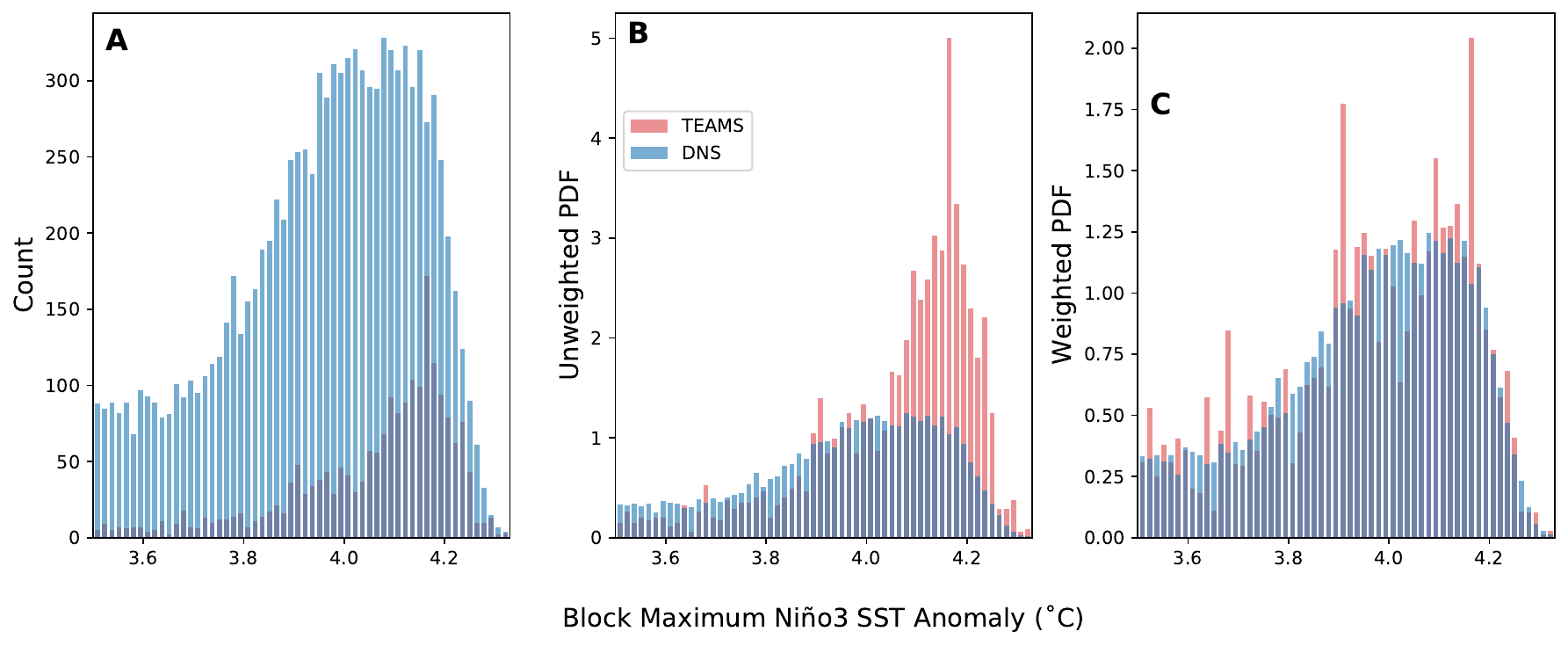}
   \end{center}
   \caption{Distributions of block maximum NINO3 SSTs (block size $T= 25$ years). Blue: DNS, $5\times10^4$ years in length. Red: TEAMS, $1.2 \times10^4$ years in length. (A) counts of extreme El Niño block maxima. Note that the TEAMS cost, and therefore count, is five times smaller than the DNS. (B) Unweighted PDF of extreme block maxima, showing TEAMS preferably generates high block maxima. (C) Weighted TEAMS PDF, matching the DNS fairly closely, particularly at extremely high values.}
   \label{fig:PDFs}
\end{figure}

Figure~\ref{fig:PDFs} shows the PDFs of block maxima calculated by TEAMS and the DNS. In general, TEAMS performed as expected: despite its smaller ensemble size (panel A), it preferentially increased the occurrence of extreme block maxima by boosting event amplitudes (panel B), while the PDF it produces given the weighting (Methods) provides a good match to the DNS, particularly in the upper tail (panel C). For the selected set of TEAMS parameters, thresholds were exceeded by child simulations with a success rate of $\sim$77\%, averaged across all TEAMS runs. TEAMS increased the overall occurrence of extreme events (block maxima $>4$ \textdegree C), shifting the PDF toward the right and creating a pronounced peak in density of events around 4.2 \textdegree C. The probability of events at the lower end of the distribution was also satisfactorily reproduced by TEAMS (Fig.~\ref{fig:PDFs}C). 

\begin{figure}[!tp]
    \begin{center}
    \includegraphics[width=0.7\textwidth]{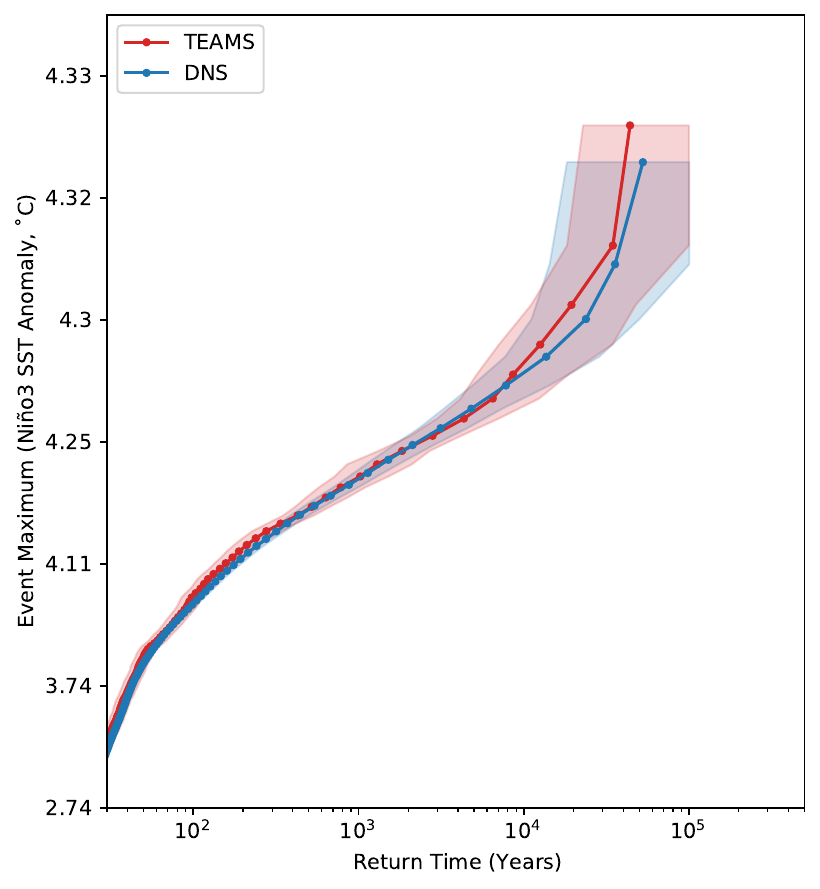}
    \end{center}
    \caption{El Niño event maximum as a function of event return time for extreme El Niño events. Event maximum is defined as the maximum monthly averaged NINO3 SST anomaly in a time block of length $T = 25$ years. The $y$-axis is stretched as in Fig.~\ref{fig:block-sizes-comparison}. Blue: DNS, with a duration of $5\times10^5$ years. Red: TEAMS-generated ensemble, with a total cost of $10^5$ years. Shading represents bootstrapped 95\% confidence intervals.} 
    \label{fig:anomaly_return_time}
    \end{figure}

TEAMS return time estimates agreed with the DNS up to return times of 60,000 years (Fig.~\ref{fig:anomaly_return_time}). Disagreement between the two return time estimates fell within the 95\% confidence intervals for the full return time range. Although the TEAMS ensemble was only 1/5 the cost of the DNS in terms of model years, the 95\% confidence intervals, particularly for longer return times, were similar between TEAMS and the DNS. Thus, TEAMS generated return time estimates that were satisfactorily accurate as compared to the model ground truth.

\subsection{TEAMS Efficiency}
\label{sec:TEAMS-efficiency}

While TEAMS is as accurate as the long DNS, it is 5 times more efficient in our application. In attempting to optimize TEAMS' efficiency, we observe diminishing returns (in terms of NINO3 increases) on increasing the maximum allowed computational cost $T_\mathrm{max}$, defined as the number of model years (including ancestors, successful children, and failed children) in a given run of TEAMS. We find that there is a sublinear relationship between the computational cost and maximum NINO3 SST anomaly achieved (Fig.~\ref{fig:max_v_cost}). In the first 3--4 rounds of splitting, each level raising (averaging of the lowest two scores from the ``alive'' ensemble members) results in a steep threshold increase. Then, in subsequent generations, the thresholds plateau as they approach a NINO3 SST value of about 4.2 \textdegree C. As a result, increasing the computational budget does not result in significant threshold increases, meaning we cannot produce events of arbitrarily large extremity by simply increasing $T_\text{max}$. Note that the apparent bound could reflect an intrinsic model property, namely a negative shape parameter in the extreme value theory formalism \cite{Coles-2001:introduction}. The shape of the far tail is a notorious source of uncertainty in climate risk analysis, and whether the plateau is present in realistic GCMs or the observed El Niños, rare event sampling offers a way to find out more efficiently. 

Increasing the number of ancestors $N$ also does not increase efficiency. This tends to concentrate more levels at the low end of the distribution and leave less budget for exploring the high end, where our main interest lies. Lastly, we found increasing the number of TEAMS iterations $M$ to be effective at decreasing the variance in TEAMS return time estimates, but we stopped once reaching our desired error range. Increasing $M$ even further did improve the accuracy (quantified by the agreement with the DNS) for the most extreme events (return times $>10,000$ years), but with an obvious accompanying decrease in efficiency. 

Overall, we find that TEAMS efficiency was most improved by adjusting the fraction of simulations killed in each level-raising $k$ and the number of repetitions $M$. Killing a larger fraction of simulations in each round more quickly pushed the threshold toward its upper bound. However, this led to a small number of samples of events for intermediate return times, degrading return time estimates for events in the 100--1,000 year range. Balancing all these ``population control'' hyperparameters is still in the realm of trial and error, and it will take more experiments on diverse systems to distill general best practices.

\subsection{An example of the development of a successful child}
\label{sec:successful-child}

\begin{figure}[!tbp]
    \begin{center}
    \includegraphics[width=0.9\textwidth]{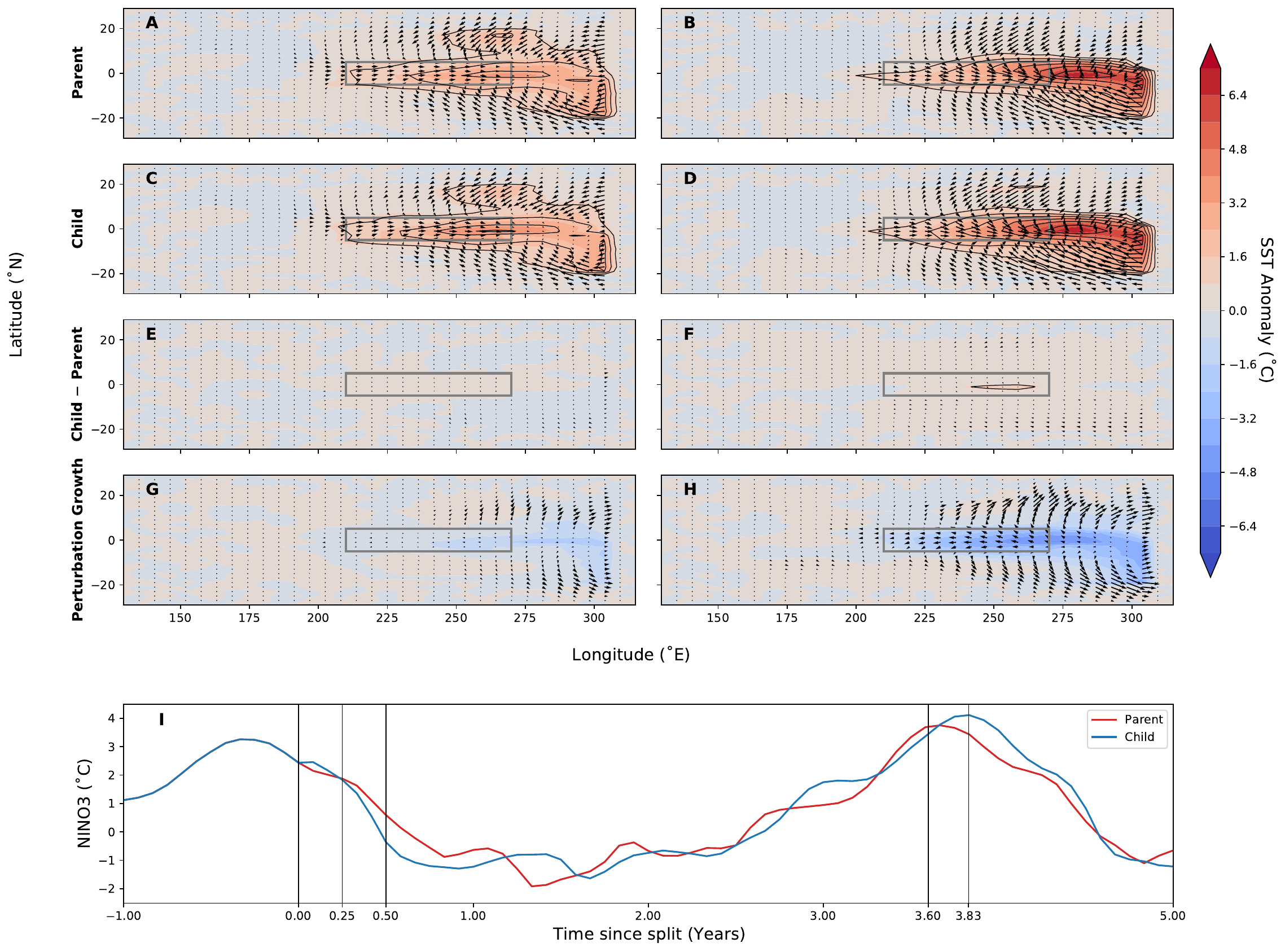}
    \end{center}
    \caption{Example of boosting an ensemble member. (A) SST anomaly field of a parent simulation at a time in which a child is split from it. (B) The state of the same parent simulation $3.60$ years later, once the threshold for its iteration of TEAMS (3.70 °C in this case) has been exceeded. (C) State of a descendant (child) of the simulation shown in A and B, immediately following its split (same time as in A), and including a small random perturbation to its SST initial conditions. (D) The state of the descendant simulation $3.83$ years after the split, when an incremented threshold of 4.08 °C has been exceeded. (E, F) are the differences of child minus parent for A and C, and for B and D. Blue and red shading represent negative and positive SST temperature anomalies. Black contour lines represent the $+1$ °C temperature isotherm. Arrows represent surface wind stress. (G,H) Show the state of the child simulation minus its initial conditions 3 months and 6 months after the split, showing that a La Niña event develops first. (I) Time series of NINO3 for the parent (red) and child simulations, starting 1 year before the split (that is, before the times of panels A, C) to 5 years after the time shown in panels B and D.}
    \label{fig:sst_composite}
\end{figure}

\begin{figure}[!tb]
    \begin{center}
    \includegraphics[width=0.6\textwidth]{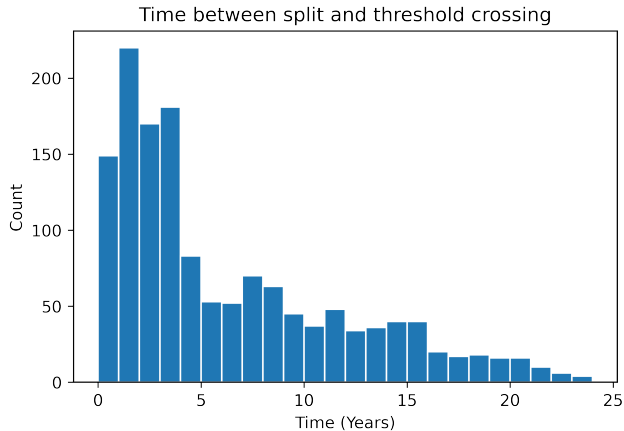}
    \end{center}
    \caption{The distribution of times between when a child simulation splits from its parent and when the child crosses its prescribed threshold.}
    \label{fig:crossing-time-distribution}
\end{figure}

To demonstrate how an initial random perturbation can make an event more extreme in TEAMS, we compare the development of a parent simulation and a boosted, more extreme child simulation (Fig.~\ref{fig:sst_composite}). Panel A shows the parent simulation at the split time, denoted $t=0$. Panel B shows the parent at its peak time, $t=3.60$ years. Panel C shows the initial conditions for the child simulation at $t=0$, different from panel A only by a small random perturbation. Panel E shows this random noise, while panel F shows the difference between the child and parent at their respective peak times: $\mathrm{child}(t=3.83)-\mathrm{parent}(t=3.60)$. In this case the child crosses its prescribed threshold of 4.08 °C at $t=3.83$ years. The parent threshold was 3.70 °C and was crossed at t=7.5 years—which sets the split time—and also at $t=3.60$ years, as shown in the figure.

By the time of their corresponding threshold crossings, both the child and parent simulations show a well-developed El Niño event, with asymmetric warming about the equator and strong winds that support this SST, according to the Bjerknes feedback. The child simulation features a slightly more pronounced peak warming (panel F), indicating that the initial perturbation achieved its goal.  Initially, though, the perturbation leads to a La Niña event, as shown by panels G and H, showing the differences between the child run at 3 and 6 months after the initial conditions and its initial state (panel C). It is interesting that the El Niño event is eventually boosted by changing initial conditions that allow a gradual buildup via a cold event first. One expects the perturbation to the initial conditions to be affected by non-normal transient growth \cite{Farrell-1988:optimal, Farrell-Ioannou-1996:generalized} shown to play a dominant role in the growth of El Niño events, whether ENSO is a damped oscillation \cite{Penland-1996:stochastic, Penland-Sardeshmukh-1995:optimal, Moore-Kleeman-1996:dynamics, Moore-Kleeman-1997:enso-I, Moore-Kleeman-1997:enso-II}, or even self-sustained and chaotic \cite{Samelson-Tziperman-2001:predictability}.

We split child simulations 7.5 years before the time the parent crosses its threshold, based on the ensemble spread time scale we calculated above. Fig.~\ref{fig:crossing-time-distribution} shows the distribution of times it takes a child simulation to cross its threshold. The peak of the distribution is at less than 5 years, with the time difference in the majority of cases being in the range of 1--4 years. This result led to our choice of an example shown in Fig.~\ref{fig:sst_composite}, where the child simulation crosses its threshold after $t = 3.83$ years. Fig.~\ref{fig:crossing-time-distribution}  shows that this early threshold crossing is not uncommon, and this reveals potential for improving the algorithm. Because the time lag is on the same order as the length of the ENSO cycle, the child simulation can sometimes cross the threshold a full cycle before its parent if that previous cycle was almost an exceedance itself. This can cause the split time to move backward in later TEAMS generations, causing the algorithm to focus on progressively earlier peaks and possibly not allow for sufficient integration time for advance splitting. This can also limit the potential for repeated boosting of a single peak, which can be more effective, especially if deterministic optimization techniques are employed where the next perturbation can learn from the previous one. In future work, it will be useful to consider alternative ways of managing the split times and perturbation structures and gain the most possible information from each new child simulation.

\section{Conclusions}

This study explored the potential for a rare event sampling technique, TEAMS \cite{Finkel-OGorman-2024:bringing,Finkel-OGorman-2025:rare}, to produce a larger amount of model data of extreme El Niño events at a lower computational cost than would be needed using a direct numerical simulation. We found that TEAMS correctly reproduces the distribution of extreme events in the intermediate complexity Cane-Zebiak model \cite{Zebiak-Cane-1987:model}, at a cost that is significantly lower than that of a DNS arriving at the same return time distribution and with the same error estimate. We explored the role of different TEAMS hyperparameters in the context of ENSO and expect the lessons will be helpful in applying the algorithm to ENSO events in more realistic and  more computationally expensive global climate models. 

It is possible that the efficiency of the TEAMS return times calculation could be improved by further exploring the TEAMS hyperparameters using a grid search or Bayesian optimization. Despite the room for further improvement, TEAMS already delivers efficiency gains. With current uncertainties about the effects of anthropogenic climate change on the frequency and intensity of the most extreme El Niño events \cite{Cai-Borlace-Lengaigne-et-al-2014:increasing}, constraining extreme El Niño event statistics is clearly a priority for climate modelers in the coming decades. TEAMS and similar rare sampling techniques have a strong potential to shed light on the drivers and underlying dynamics of extreme El Niño events in the present and future warmer climates.

\section*{Conflict of Interest}

The authors declare no conflict of interest.

\section*{Open Research Section}

N/A

\acknowledgments
DA was funded in part by the US National Science Foundation through award NSF RISE–2425898. Support was provided for JF by Schmidt Sciences, LLC during the time of this research. ET was funded by the Department of Energy (DOE) Office of Science Biological and Environmental Research grant DE-SC0023134 and the Harvard Dean’s Competitive Fund for Promising Scholarship, and thanks the Weizmann Institute of Science for its hospitality during parts of this work.

\newpage
\bibliography{export}

@String{JAMES           =       "Journal of Advances in Modeling Earth Systems"}

@String{JAS		=	"Journal of the Atmospheric Sciences"}

@String{JOC		=	"Journal of Climate"}

@String{MWR		=	"Monthly Weather Review"}

@String{NATURE		=	"Nature"}

@article{Abbot-Webber-Hadden-et-al-2021:rare,
  title =	 {Rare event sampling improves {Mercury} instability
                  statistics},
  volume =	 923,
  issn =	 {0004-637X},
  doi =		 {10.3847/1538-4357/ac2fa8},
  language =	 {en},
  number =	 2,
  urldate =	 {2024-11-12},
  journal =	 {The Astrophysical Journal},
  author =	 {Abbot, Dorian S. and Webber, Robert J. and Hadden,
                  Sam and Seligman, Darryl and Weare, Jonathan},
  month =	 dec,
  year =	 2021,
  pages =	 236,
}

@incolleCtion{Bucklew-2004:importance,
  address =	 {New York, NY},
  title =	 {Importance {Sampling}},
  isbn =	 {978-1-4757-4078-3},
  url =		 {10.1007/978-1-4757-4078-3\_4},
  language =	 {en},
  urldate =	 {2024-11-12},
  booktitle =	 {Introduction to {Rare} {Event} {Simulation}},
  publisher =	 {Springer},
  author =	 {Bucklew, James Antonio},
  editor =	 {Bucklew, James Antonio},
  year =	 2004,
  pages =	 {57--73}
}

@article{Cai-Borlace-Lengaigne-et-al-2014:increasing,
  title =	 {Increasing frequency of extreme El Ni{\~n}o events
                  due to greenhouse warming},
  author =	 {Cai, Wenju and Borlace, Simon and Lengaigne,
                  Matthieu and Van Rensch, Peter and Collins, Mat and
                  Vecchi, Gabriel and Timmermann, Axel and Santoso,
                  Agus and McPhaden, Michael J and Wu, Lixin and
                  England, MH and Wang, G and Guilyardi, E and Jin,
                  F-F},
  journal =	 {Nature Climate Change},
  volume =	 4,
  number =	 2,
  pages =	 {111--116},
  year =	 2014,
  publisher =	 {Nature Publishing Group}
}

@article{Cai-Kalnay-Toth-2003:bred,
  title =	 {Bred Vectors of the ZebiakCane Model and Their
                  Potential Application to ENSO Predictions},
  author =	 {Cai, M. and E. Kalnay and Z. Toth},
  journal =	 JOC,
  volume =	 16,
  pages =	 {40-56},
  year =	 2003
}

@Article{Cane-2005:evolution,
  author =	 {M. A. Cane},
  title =	 {The evolution of {El Ni\~no}, past and future},
  journal =	 {Earth and Planetary Science Letters},
  year =	 2005,
  volume =	 230,
  pages =	 {227--240}
}

@article{Clement-Seager-Cane-2000:suppression,
  title =	 {Suppression of {El} {Niño} during the
                  {Mid}-{Holocene} by changes in the {Earth}'s orbit},
  volume =	 15,
  copyright =	 {Copyright 2000 by the American Geophysical Union.},
  issn =	 {1944-9186},
  url =
                  {https://onlinelibrary.wiley.com/doi/abs/10.1029/1999PA000466},
  doi =		 {10.1029/1999PA000466},
  language =	 {en},
  number =	 6,
  urldate =	 {2024-11-14},
  journal =	 {Paleoceanography},
  author =	 {Clement, Amy C. and Seager, Richard and Cane, Mark
                  A.},
  year =	 2000,
  pages =	 {731--737},
}

@article{Farrell-1988:optimal,
  author =	 "Brian Farrell",
  title =	 "Optimal excitation of neutral {Rossby} waves",
  journal =	 JAS,
  year =	 1988,
  volume =	 45,
  pages =	 "163-172"
}

@article{Farrell-Ioannou-1996:generalized,
  author =	 "Farrell, B. F. and P. J. Ioannou",
  year =	 1996,
  title =	 "Generalized stability theory part {I}: autonomous
                  operators",
  journal =	 JAS,
  volume =	 53,
  pages =	 "2025-2040"
}

@article{Finkel-OGorman-2024:bringing,
  title =	 {Bringing statistics to storylines: {Rare} event
                  sampling for sudden, transient extreme events},
  volume =	 16,
  issn =	 {1942-2466},
  shorttitle =	 {Bringing {Statistics} to {Storylines}},
  url =
                  {https://onlinelibrary.wiley.com/doi/abs/10.1029/2024MS004264},
  doi =		 {10.1029/2024MS004264},
  language =	 {en},
  number =	 6,
  urldate =	 {2024-10-04},
  journal =	 {Journal of Advances in Modeling Earth Systems},
  author =	 {Finkel, Justin and O'Gorman, Paul A.},
  year =	 2024,
}

@article{Finkel-OGorman-2025:rare,
  title =	 {Rare event sampling for moving targets: extremes of
                  temperature and daily precipitation in a general
                  circulation model},
  author =	 {Justin Finkel and Paul A. O'Gorman},
  journal =	 {ArXiv},
  year =	 2025,
  eprint =	 {2508.13120},
  archivePrefix ={arXiv},
  primaryClass = {physics.ao-ph},
  url =		 {https://arxiv.org/abs/2508.13120},
}

@article{Geng-Jin-2022:enso,
  title =	 {{ENSO} {Diversity} {Simulated} in a {Revised}
                  {Cane}-{Zebiak} {Model}},
  volume =	 10,
  issn =	 {2296-6463},
  doi =		 {10.3389/feart.2022.899323},
  language =	 {English},
  urldate =	 {2025-01-23},
  journal =	 {Frontiers in Earth Science},
  author =	 {Geng, Licheng and Jin, Fei-Fei},
  month =	 apr,
  year =	 2022,
}

@article{Jansen-Mandjes-Taimre-2023:rare,
  title =	 {Rare-event simulation techniques for structured
                  fisheries models},
  volume =	 28,
  issn =	 {1573-2967},
  doi =		 {10.1007/s10666-023-09900-6},
  language =	 {en},
  number =	 5,
  urldate =	 {2024-11-12},
  journal =	 {Environmental Modeling \& Assessment},
  author =	 {Jansen, Hermanus M. and Mandjes, Michel and Taimre,
                  Thomas},
  month =	 oct,
  year =	 2023,
  pages =	 {907--924},
}

@article{Kahn-Harris-1951:estimation,
  title =	 {Estimation of particle transmission by random
                  sampling},
  volume =	 12,
  journal =	 {National Bureau of Standards Applied Mathematics
                  Series},
  author =	 {Kahn, Herman and Harris, Theodore E.},
  year =	 1951,
  pages =	 {27--30},
}

@article{Lestang-Ragone-Bréhier-et-al-2018:computing,
  title =	 {Computing return times or return periods with rare
                  event algorithms},
  volume =	 2018,
  issn =	 {1742-5468},
  url =		 {https://dx.doi.org/10.1088/1742-5468/aab856},
  doi =		 {10.1088/1742-5468/aab856},
  language =	 {en},
  number =	 4,
  urldate =	 {2024-10-04},
  journal =	 {Journal of Statistical Mechanics: Theory and
                  Experiment},
  author =	 {Lestang, Thibault and Ragone, Francesco and Bréhier,
                  Charles-Edouard and Herbert, Corentin and Bouchet,
                  Freddy},
  month =	 apr,
  year =	 2018,
  pages =	 {43-213},
}

@article{Lucente-Herbert-Bouchet-2022:committor,
  title =	 {Committor functions for climate phenomena at the
                  predictability margin: {The} example of {El}
                  {Niño}–{Southern} {Oscillation} in the {Jin} and
                  {Timmermann} {Model}},
  volume =	 79,
  shorttitle =	 {Committor {Functions} for {Climate} {Phenomena} at
                  the {Predictability} {Margin}},
  doi =		 {10.1175/JAS-D-22-0038.1},
  language =	 {en},
  number =	 9,
  urldate =	 {2024-11-12},
  journal =	 {Journal of the Atmospheric Sciences},
  author =	 {Lucente, Dario and Herbert, Corentin and Bouchet,
                  Freddy},
  month =	 sep,
  year =	 2022,
  pages =	 {2387--2400},
}

@article{Mann-Cane-Zebiak-et-al-2005:volcanic,
  author =	 {Mann, M. E. and Cane, M. A. and Zebiak, S. E. and
                  Clement, A.},
  title =	 {Volcanic and solar forcing of the tropical Pacific
                  over the past 1000 years},
  journal =	 JOC,
  year =	 2005,
  volume =	 18,
  number =	 3,
  pages =	 {447-456},
  month =	 {Feb.}
}

@article{McPhaden-1999:genesis,
  title =	 {Genesis and evolution of the 1997-98 El Ni{\~n}o},
  author =	 {McPhaden, Michael J},
  journal =	 {Science},
  volume =	 283,
  number =	 5404,
  pages =	 {950--954},
  year =	 1999,
  publisher =	 {American Association for the Advancement of Science}
}

@article{McPhaden-Zebiak-Glantz-2006:enso,
  title =	 {ENSO as an integrating concept in earth science},
  author =	 {McPhaden, Michael J and Zebiak, Stephen E and
                  Glantz, Michael H},
  journal =	 {science},
  volume =	 314,
  number =	 5806,
  pages =	 {1740--1745},
  year =	 2006,
  publisher =	 {American Association for the Advancement of Science}
}

@article{Moore-Kleeman-1996:dynamics,
  author =	 {Moore, A. M. and R. Kleeman},
  title =	 {The dynamics of error growth and predictability in a
                  coupled model of {ENSO}},
  journal =	 {Q. J. R. Meteor. Soc.},
  year =	 1996,
  volume =	 122,
  pages =	 {1405-1446},
}

@article{Moore-Kleeman-1997:enso-I,
  author =	 {Moore, A. M. and R. Kleeman},
  title =	 {The singular vectors of a coupled ocean-atmosphere
                  model of {ENSO}, {I}, Thermodynamics, energetics and
                  error growth},
  journal =	 {Q. J. R. Meteor. Soc.},
  year =	 1997,
  volume =	 123,
  pages =	 {953-981},
}

@article{Moore-Kleeman-1997:enso-II,
  author =	 {Moore, A. M. and R. Kleeman},
  title =	 {The singular vectors of a coupled ocean-atmosphere
                  model of {ENSO, II}, Sensitivity studies and
                  dynamical interpretation.},
  journal =	 {Q. J. R. Meteor. Soc.},
  year =	 1997,
  volume =	 123,
  pages =	 {983-1006},
}

@article{Penland-1996:stochastic,
  author =	 {C. Penland},
  title =	 {A stochastic model of IndoPacific sea surface
                  temperature anomalies},
  journal =	 {Physica D},
  volume =	 98,
  number =	 {2-4},
  pages =	 {534-558},
  month =	 {nov 15},
  year =	 1996
}

@article{Penland-Sardeshmukh-1995:optimal,
  author =	 {C. Penland and P. D. Sardeshmukh},
  title =	 {The optimal-growth of tropical sea-surface
                  temperature anomalies},
  journal =	 JOC,
  volume =	 8,
  number =	 8,
  pages =	 {1999-2024},
  month =	 aug,
  year =	 1995
}

@article{Ragone-Wouters-Bouchet-2018:computation,
  title =	 {Computation of extreme heat waves in climate models
                  using a large deviation algorithm},
  volume =	 115,
  url =		 {https://www.pnas.org/doi/10.1073/pnas.1712645115},
  doi =		 {10.1073/pnas.1712645115},
  number =	 1,
  urldate =	 {2024-11-12},
  journal =	 {Proceedings of the National Academy of Sciences},
  author =	 {Ragone, Francesco and Wouters, Jeroen and Bouchet,
                  Freddy},
  month =	 jan,
  year =	 2018,
  pages =	 {24--29},
}

@article{Webber-Plotkin-ONeill-et-al-2019:practical,
  title =	 {Practical rare event sampling for extreme mesoscale
                  weather},
  volume =	 29,
  issn =	 {1054-1500},
  url =		 {https://doi.org/10.1063/1.5081461},
  doi =		 {10.1063/1.5081461},
  number =	 5,
  urldate =	 {2024-10-28},
  journal =	 {Chaos: An Interdisciplinary Journal of Nonlinear
                  Science},
  author =	 {Webber, Robert J. and Plotkin, David A. and ONeill,
                  Morgan E and Abbot, Dorian S. and Weare, Jonathan},
  month =	 may,
  year =	 2019,
  pages =	 053109,
}

@book{Wilkinson-2004:status,
  title =	 {{Status of Coral Reefs of the World}: 2004},
  volume =	 2,
  publisher =	 {Australian Institute of Marine Science},
  author =	 {Wilkinson, Clive R.},
  year =	 2004
}

@article{Xie-Jin-2018:two,
  title =	 {Two leading ENSO modes and El Ni{\~n}o types in the
                  Zebiak--Cane model},
  author =	 {Xie, Ruihuang and Jin, Fei-Fei},
  journal =	 {Journal of Climate},
  volume =	 31,
  number =	 5,
  pages =	 {1943--1962},
  year =	 2018
}

@article{Zebiak-Cane-1987:model,
  author =	 "Zebiak, S. E. and M. A. Cane",
  title =	 "{A model {El Ni{\~n}o-Southern Oscillation}}",
  journal =	 MWR,
  volume =	 115,
  pages =	 "2262-2278",
  year =	 1987
}

@article{Zuckerman-Chong-2017:weighted,
  title =	 {Weighted ensemble simulation: {Review} of
                  methodology, applications, and software},
  volume =	 46,
  issn =	 {1936-122X, 1936-1238},
  shorttitle =	 {Weighted {Ensemble} {Simulation}},
  url =
                  {https://www-annualreviews-org.ezp-prod1.hul.harvard.edu/content/journals/10.1146/annurev-biophys-070816-033834},
  doi =		 {10.1146/annurev-biophys-070816-033834},
  language =	 {en},
  urldate =	 {2024-11-12},
  journal =	 {Annual Review of Biophysics},
  author =	 {Zuckerman, Daniel M. and Chong, Lillian T.},
  month =	 may,
  year =	 2017,
  pages =	 {43--57},
}

@Article{Jacques-Dumas-Westen-Bouchet-et-al-2022:data,
  AUTHOR =	 {Jacques-Dumas, V. and van Westen, R. M. and Bouchet,
                  F. and Dijkstra, H. A.},
  TITLE =	 {Data-driven methods to estimate the committor
                  function in conceptual ocean models},
  JOURNAL =	 {EGUsphere},
  VOLUME =	 2022,
  YEAR =	 2022,
  PAGES =	 {1--35},
  URL =
                  {https://egusphere.copernicus.org/preprints/egusphere-2022-1362/},
  DOI =		 {10.5194/egusphere-2022-1362}
}

@article{Finkel-Webber-Gerber-et-al-2021:learning,
  author =	 "Justin Finkel and Robert J. Webber and Edwin P.
                  Gerber and Dorian S. Abbot and Jonathan Weare",
  title =	 "Learning Forecasts of Rare Stratospheric Transitions
                  from Short Simulations",
  journal =	 "Monthly Weather Review",
  year =	 2021,
  publisher =	 "American Meteorological Society",
  address =	 "Boston MA, USA",
  volume =	 149,
  number =	 11,
  doi =		 "10.1175/MWR-D-21-0024.1",
  pages =	 "3647 - 3669",
  url =
                  "https://journals.ametsoc.org/view/journals/mwre/149/11/MWR-D-21-0024.1.xml"
}

@article {Finkel-Abbot-Weare-2020:path,
  author =	 "Justin Finkel and Dorian S. Abbot and Jonathan
                  Weare",
  title =	 "Path Properties of Atmospheric Transitions:
                  Illustration with a Low-Order Sudden Stratospheric
                  Warming Model",
  journal =	 "Journal of the Atmospheric Sciences",
  year =	 2020,
  publisher =	 "American Meteorological Society",
  address =	 "Boston MA, USA",
  volume =	 77,
  number =	 7,
  doi =		 "10.1175/JAS-D-19-0278.1",
  pages =	 "2327 - 2347",
  url =
                  "https://journals.ametsoc.org/view/journals/atsc/77/7/jasD190278.xml"
}

@article{Wouters-Schiemann-Shaffrey-2023:rare,
  author =	 {Wouters, Jeroen and Schiemann, Reinhard K. H. and
                  Shaffrey, Len C.},
  title =	 {Rare Event Simulation of Extreme European Winter
                  Rainfall in an Intermediate Complexity Climate
                  Model},
  journal =	 {Journal of Advances in Modeling Earth Systems},
  volume =	 15,
  number =	 4,
  pages =	 {e2022MS003537},
  doi =		 {https://doi.org/10.1029/2022MS003537},
  url =
                  {https://agupubs.onlinelibrary.wiley.com/doi/abs/10.1029/2022MS003537},
  eprint =
                  {https://agupubs.onlinelibrary.wiley.com/doi/pdf/10.1029/2022MS003537},
  note =	 {e2022MS003537 2022MS003537},
  year =	 2023
}

@article{Plotkin-Webber-ONeill-et-al-2019:maximizing,
  title =	 {Maximizing simulated tropical cyclone intensity with
                  action minimization},
  author =	 {Plotkin, David A and Webber, Robert J and O'Neill,
                  Morgan E and Weare, Jonathan and Abbot, Dorian S},
  journal =	 {Journal of Advances in Modeling Earth Systems},
  volume =	 11,
  number =	 4,
  pages =	 {863--891},
  year =	 2019,
  publisher =	 {Wiley Online Library}
}

@book{Coles-2001:introduction,
  title =	 {An introduction to statistical modeling of extreme
                  values},
  author =	 {Coles, Stuart},
  series =	 {Springer Series in Statistics},
  year =	 2001,
  publisher =	 {Springer},
  doi =		 {10.1007/978-1-4471-3675-0},
  isbn =	 {978-1-85233-459-8},
  issn =	 {0172-7397},
  edition =	 1,
}

@article{Tziperman-Cane-Zebiak-1995:irregularity,
  author =	 {E. Tziperman and M. A. Cane and S. E. Zebiak},
  title =	 {Irregularity and locking to the seasonal cycle in an
                  {ENSO} prediction model as explained by the
                  quasi-periodicity route to chaos},
  journal =	 JAS,
  volume =	 52,
  number =	 3,
  pages =	 {293-306},
  month =	 feb,
  year =	 1995,
  doi =		 {10.1175/1520-0469(1995)052$<$0293:IALTTS$>$2.0.CO;2},
  link =	 "\href{http://www.seas.harvard.edu/climate/eli/reprints/Tziperman-Cane-Zebiak-1995.pdf}{download}"
}

@article{Tziperman-Zebiak-Cane-1997:mechanisms,
  author =	 {E. Tziperman and S. E. Zebiak and M. A. Cane},
  title =	 {Mechanisms of seasonal - {ENSO} interaction},
  journal =	 JAS,
  volume =	 54,
  number =	 1,
  pages =	 {61-71},
  month =	 jan,
  year =	 1997,
  doi =		 {10.1175/1520-0469(1997)054$<$0061:MOSEI$>$2.0.CO;2},
  link =	 "\href{http://www.seas.harvard.edu/climate/eli/reprints/Tziperman-Zebiak-Cane-1997.pdf}{download}"
}

@Article{Samelson-Tziperman-2001:predictability,
  author =	 {Samelson, R. and E. Tziperman},
  title =	 {Instability of the chaotic {ENSO}: {The}
                  growth-phase predictability barrier},
  journal =	 JAS,
  year =	 2001,
  doi =		 {10.1175/1520-0469(2001)058$<$3613:IOTCET$>$2.0.CO;2},
  volume =	 58,
  pages =	 {3613-3625},
  link =
                  "\href{http://www.seas.harvard.edu/climate/eli/reprints/Samelson-Tziperman-2001.pdf}{download}"
}

@article{Noyelle2025evolution,
author = {Noyelle, Robin and Caubel, Arnaud and Meurdesoif, Yann and Faranda, Davide and Yiou, Pascal},
title = {Evolution of the Dynamics of Centennial Hot Summers in Western Europe With Climate Change},
journal = {Geophysical Research Letters},
volume = {52},
number = {14},
pages = {e2025GL115552},
doi = {https://doi.org/10.1029/2025GL115552},
url = {https://agupubs.onlinelibrary.wiley.com/doi/abs/10.1029/2025GL115552},
eprint = {https://agupubs.onlinelibrary.wiley.com/doi/pdf/10.1029/2025GL115552},
note = {e2025GL115552 2025GL115552},
year = {2025}
}

@article{Cerou2019adaptive,
author = {Cérou,Frédéric  and Guyader,Arnaud  and Rousset,Mathias },
title = {Adaptive multilevel splitting: Historical perspective and recent results},
journal = {Chaos: An Interdisciplinary Journal of Nonlinear Science},
volume = {29},
number = {4},
pages = {043108},
year = {2019},
doi = {10.1063/1.5082247},

URL = {
        https://doi.org/10.1063/1.5082247

},
eprint = {
        https://doi.org/10.1063/1.5082247

}

}

@article{lancelin2025ai,
  title =	 {AI-boosted rare event sampling to characterize
                  extreme weather},
  author =	 {Lancelin, Amaury and Wikner, Alex and Dubus, Laurent
                  and Priol, Cl{\'e}ment Le and Abbot, Dorian S and
                  Bouchet, Freddy and Hassanzadeh, Pedram and Weare,
                  Jonathan},
  journal =	 {arXiv preprint arXiv:2510.27066},
  year =	 {2025}
}

\end{document}